\documentclass[a4paper,11pt]{article}
\pdfoutput=1 

\usepackage{jheppub} 

\usepackage[T1]{fontenc} 
\usepackage{float}

\title{Supersymmetric traversable wormholes}
\author[a]{Andr\'{e}s Anabal\'{o}n}
\author[b,c]{Bernard de Wit}
\author[d]{Julio Oliva}

\affiliation[a]{Departamento de Ciencias, Facultad de Artes Liberales,
Universidad Adolfo Ib\'{a}$\tilde{\text{n}}$ez, Avenida Padre Hurtado
750, Vi$\tilde{n}$a del Mar, Chile}
\affiliation[b]{Institute for Theoretical Physics, Utrecht University,
  Princetonplein 5, 3584 CC Utrecht, The Netherlands}
\affiliation[c]{Nikhef Theory Group, Science Park 105, 1098 XG Amsterdam,
The   Netherlands}
\affiliation[d]{Departamento de F\'{\i}sica, Universidad de Concepci\'{o}n, Casilla
  160-C, Concepci\'{o}n, Chile.}

\emailAdd{andres.anabalon@uai.cl} \emailAdd{B.deWit@uu.nl}
\emailAdd{juoliva@udec.cl} \abstract{ We study traversable wormhole
  solutions in pure gauged $N\!=\!2$ supergravity with and without
  electromagnetic fields, which are locally isometric under
  $\mathrm{SO}(2,1)\!\times\!\mathrm{SO}(1,1)$. The model allows for
  1/2-BPS wormhole solutions whose corresponding globally defined
  Killing spinors are presented. A non-contractible cycle can be
  obtained by compactifying one of the coordinates which leaves the
  residual supersymmetry unaffected, the isometry group is now globally $\mathrm{SO}(2,1)\!\times\!\mathrm{SO}(2)$. The wormholes connect two asymptotic,
  locally $\mathrm{AdS}_4$ regions and depend on certain electric and
  magnetic charge parameters and, implicitly, on the range of the
  compact coordinate around the throat. We provide an analysis of the
  boundary of the spacetime and show that it can be either
  disconnected or not, depending on the values of the parameters in the metric.  Finally, we show how that these space-times avoid a
  topological censorship theorem.}
\begin{document}

\maketitle

\section{Introduction}

\label{introduction} 
The term ``wormhole'' was first introduced in a paper by Fuller and Wheeler 
\cite{Fuller:1962zza}, where credit was given to Weyl for the idea of having
a non-simply connected space-time. Indeed, this is an essential feature of a
wormhole has. It is the mathematical realization of the idea that within the
same universe it is possible to travel between two points using two
different paths. These ideas were thought to happen within a single universe
with a single boundary. In this paper we will explicitly construct such a
non-simply connected space-time with a single boundary by compactifying one
non-compact coordinate of global $AdS_4$.

We should add that the idea of a wormhole is also associated with the work
by Einstein and Rosen \cite{Einstein:1935tc}, where non-singular coordinate
patches of the Schwarzschild and the Reissner-Nordstr\"{o}m black holes were
studied. However, it is important to realize that the latter wormholes are
very different from the one of Fuller and Wheeler, because the
Einstein-Rosen bridge connects two disjoint universes that are causally
disconnected. In this paper, we show that when our single boundary wormhole
is extended to the non-constant-curvature case a parallel propagating
singularity develops at the boundary.

Despite these interesting features, wormholes have been widely regarded as a
science fiction character. As discussed in \cite{Morris:1988tu,
Morris:1988cz}, this is due to the fact that the null-energy condition has
to be violated at the throat of a spherically symmetric, static wormhole. 
\footnote{%
It is interesting to note that there is a  widespread belief in the
literature against the existence of  traversable wormholes under physically
sensible energy  conditions. This is primarily based on the analysis of \cite%
{Hochberg:1998ii}, where it is claimed that a four-dimensional  space-time
cannot have a wormhole with a minimal $\mathbb{S}^{2}$  when the null-energy
condition holds. However, the wormhole studied  in this paper has a minimal $%
\mathbb{S}^{1}$ and therefore the  analysis of \cite{Hochberg:1998ii} does
not apply. Indeed, the  matter content that we use does satisfy the
null-energy  condition.} Hence, for asymptotically flat space-times there is
not much hope for wormholes to exist in a physically sensible situation. The
situation changes in asymptotically AdS space-times. When the
four-dimensional space-time is Einstein, and its conformal boundary has
positive scalar curvature with a space-time that is everywhere regular, then
the boundary cannot have more than one connected component. If the boundary
has negative scalar curvature, it is possible to construct a Euclidean
wormhole by identifications in global AdS \cite{Witten:1999xp}. Some of
these identifications have been analyzed in Euclidean AdS and several
arguments have been given against the stability of these wormholes \cite%
{Maldacena:2004rf}. A standard one is that conformally coupled scalar fields
living on a conformal boundary of negative curvature will have an action
that is unbounded from below. This is indeed correct when the boundary is in
the conformal class of $\mathbb{H} ^{2}\times \mathbb{R}$. However, when the
boundary is itself an AdS space-time this argument no longer applies for
conformally coupled scalar fields, as their masses are always above the
Breitenlohner-Freedman bound \cite{Breitenlohner:1982bm,
Breitenlohner:1982jf}.

This observation motivated the work in \cite{Anabalon:2018rzq}, where a
class of geometrically non-trivial solutions was constructed where the
boundary is composed of two, possibly warped, ${\mathrm{AdS}}_{3}$
space-times. Here we show that when the boundary is not warped ${\mathrm{AdS}%
}_{3}$, but just a locally $\mathrm{AdS}_{3}$ space-time, the proper time
needed to go from one of the $\mathrm{AdS}_{3}$ components of the boundary
to the other is longer through the bulk than through the boundary. This is
what is called a long wormhole \cite{Maldacena:2018gjk}. When the boundary
has two warped $\mathrm{AdS}_3$ components, we shall argue that they are
actually disconnected due to the existence of a parallelly propagated
curvature singularity at what should be boundary of warped $\mathrm{AdS}_3$.
However, a wormhole is not characterized exclusively by the number of
boundaries it has. As discussed in \cite{Fuller:1962zza}, the wormhole can
only exist provided the manifold is \textit{non-simply connected}. When the
non-contractible cycle has minimal length, one is dealing with a wormhole
throat. As we shall see, this can already be achievable at the level of a
constant curvature manifold, namely a locally $\mathrm{AdS}_{4}$ space-time.
In the coordinates that we use this is implemented by requiring that one
coordinate of global $\mathrm{AdS}_{4}$ to be compactified. After this
identification the space-time is no longer globally $\mathrm{AdS}_{4}$ but a
Lorentzian wormhole of constant curvature. The same phenomenon exists for
the solutions discussed in \cite{Anabalon:2018rzq}, which contain
non-trivial electromagnetic fields. These solutions also describe wormholes
upon introducing a non-contractible cycle.

In this paper we discuss a number of new results regarding the solutions
proposed in \cite{Anabalon:2018rzq}. First of all we will consider these
solutions in the context of $N=2$ pure gauged supergravity, which contains a
non-trivial photon field. We will prove that they admit 1/2 BPS solutions
that respect the non-simply connected topology. The reason for the latter is
that the Killing spinors do not depend on the compactified coordinate. In
the constant-curvature case, half of the supersymmetries are no longer
globally defined, so that only half of the Killing spinors will exist.

Finally, we discuss topological censorship, which is the claim disconnected
component of the boundary can not communicate in an asymptotically locally
AdS space-time, we show how our solutions do not satisfy the hypothesis of
this quite general theorem \cite{Galloway:1999br}.\footnote{%
We thank Juan Maldacena to bring this reference to our notice and to Nava
Gaddam and Krinio Marouda for suggesting that a violation of the generic
condition is the way out of the theorem.}

The plan of the paper is as follows. Gauged $N=2$ supergravity is introduced
in section \ref{sec:the model}, followed by a discussion of the wormhole
solutions in section \ref{sec:einst-maxw-wormhole}. In section \ref%
{sec:susy-wormhole-solutions} we then consider the possibility of
supersymmetric wormholes by proving that the integrability condition for the
existence of Killing spinors is satisfied. Subsequently we present an
explicit construction of the Killing spinors in section \ref%
{sec:susy-wormhole-solutions} and show that they are globally defined and
fully compatible with the global features of the background. Some geometric
aspects of supersymmetric wormholes are discussed in section \ref%
{sec:geometric-aspects}. When the boundary of the space-time has two $%
\mathrm{AdS}_{3}$ components, we show how they can be compactified in a
single $\mathbb{S}^{2}\times \mathbb{R}$, which implies that these two
boundaries are connected. We compute the time that a massless particle needs
to go through the bulk wormhole and verify that indeed is longer than takes
a shortcut through a geodesic lying completely on the boundary. Moreover, we
show that these two geodesics are not-homotopic. Hence, the wormhole agrees
with the expectations of \cite{Maldacena:2018gjk}. When the boundary of the
space-time has two warped $\mathrm{AdS}_{3}$ components, we argue that the
existence of a parallelly propagated curvature singularity imply that the
boundaries are disconnected. Finally we discuss topological censorship for
our solutions and point out how we avoid a well-known theorem. Our
conclusions are presented in section \ref{sec:discussion}.


\section{The supergravity model}

\label{sec:the model} \setcounter{equation}{0} 
In this section we present various features of pure $N=2$ supergravity with
electrically charged gravitini. As is well known, supersymmetry will then
imply the presence of a cosmological term whose coefficient is proportional
to the square of the gravitino charge. This theory was originally
constructed in terms of the physical fields \cite%
{Freedman:1976aw,Fradkin:1976xz}, whose supersymmetry transformations only
close under commutation up to equations of motion. Subsequently two
alternative constructions were presented based on the superconformal
multiplet calculus \cite{deWit:1980lyi,deWit:1984wbb,deWit:1984rvr}. The
physical degrees of freedom of this theory are described by the vierbein
field $e_{\mu }{\!}^{a}$, electrically charged gravitini $\psi_{\mu}$, and a
photon field $A_{\mu }$. In addition we employ a spin-connection field $%
\omega _{\mu }{\!}^{ab}$ associated with (local) Lorentz transformations,
which is not an independent field. The gravitational coupling constant has
been absorbed in the fields, and the gravitino charge is equal to $q$. The
gravitino fields act as the gauge fields associated with local supersymmetry.%
\footnote{
World indices $\mu ,\nu ,\ldots $, and tangent-space indices $a,b,\ldots $,
both run from $0$ to $3$. The gamma matrices satisfy $\{\gamma _{a},\gamma
_{b}\}=2\,\eta _{ab}\,\mathbf{1}$, where the tangent-space metric equals $%
\eta_{ab}=\mathrm{diag}(-1,1,1,1)$. Furthermore $\gamma ^{5}=-\mathrm{i}%
\gamma ^{0}\gamma ^{1}\gamma ^{2}\gamma ^{3}$. In four space-time dimensions
the charge-conjugation matrix $C$ is anti-symmetric and gamma matrices $%
\gamma _{a}$ satisfy $C\gamma _{a}C^{-1}=-\gamma _{a}{\!}^ \mathrm{T}$.}

The $N=2$ supersymmetry transformations are described by two Majorana spinor
parameters distinguished by an index $i=1,2$, and are decomposed in terms of
their chiral components. The reason is that $N=2$ supersymmetry in four
space-time dimensions has a chiral R-symmetry group $\mathrm{SU}(2)\times 
\mathrm{U}(1)$. Therefore it makes sense to consider a doublet of
positive-chirality spinor parameters denoted by $\epsilon ^{i}$ and a
similar doublet of negative-chirality parameters $\epsilon _{i}$, which each
transform according under R-symmetry. As it turns out the electromagnetic
gauge transformations correspond to an abelian subgroup of the $\mathrm{SU}%
(2)$ R-symmetry group. We denote the generator of this subgroup by $t^{i}{\!}%
_{j}$, which is thus an anti-hermitian traceless matrix. The fact that we
are dealing with Majorana spinors implies that the Dirac conjugate of a
chiral spinor is proportional to the anti-chiral spinor, and vice versa. For
instance, the Dirac conjugate of $\epsilon ^{i}$ is denoted by $\bar{\epsilon%
}_{i}$, where the conjugate must carry a lower $\mathrm{SU}(2)$ index, and $%
C\,\bar{\epsilon}_{i}{\!}^{\mathrm{T}}=\epsilon _{i}$, where $C$ is the
charge-conjugation matrix and the superscripte $\mathrm{T}$ indicates that
we have taken the transpose.

Obviously these spinorial properties are carried over to the gravitino
fields, where we again distinguish two chiral doublets satisfying 
\begin{equation}  \label{eq:spinor-components}
\gamma_5\, \psi_\mu{\!}^i = + \psi_\mu{\!}^i\,,\qquad \gamma_5\,
\psi_{\mu\,i} = - \psi_{\mu\,i} \,.
\end{equation}
The results given below were taken from \cite{deWit:1984rvr}, where a large
class of $N=2$ theories was presented. Here we consider the following
supergravity Lagrangian (up to terms quartic in the gravitini), 
\begin{align}  \label{eq:Lagrangian}
\mathcal{L}=& \,-{\textstyle\frac{1}{2}}e\,R(\omega ,e)-\tfrac{1}{8}
e\,F(A)_{\mu \nu }\,F(A)^{\mu \nu }  \nonumber \\[1mm]
& -\tfrac{1}{2}e\,\big[\bar{\psi}_{\mu }{\!}^{i}\,\gamma ^{\mu \nu \rho }\, 
\mathcal{D}_{\nu }\psi _{\rho \,i}-\bar{\psi}_{\mu \,i}\,\gamma ^{\mu \nu
\rho }\,\mathcal{D}_{\nu }\psi _{\rho }{\!}^{i}\,\big]  \nonumber \\[1mm]
& +\tfrac{1}{8}F(A)^{\rho \sigma }\big[\varepsilon _{ij}\bar{\psi}_{\mu } {\!%
}^{i}\gamma ^{\lbrack \mu }\gamma _{\rho \sigma }\gamma ^{\nu ]}\psi _{\nu } 
{\!}^{j}+\varepsilon ^{ij}\bar{\psi}_{\mu \,i}\gamma ^{\lbrack \mu }\gamma
_{\rho \sigma }\gamma ^{\nu ]}\psi _{\nu \,j}\,\big]  \nonumber \\[1mm]
& \,+\tfrac{1}{2}\sqrt{2}\,qe\,\big[\varepsilon _{ik}\,t^{k}{\!}_{j}\, \bar{%
\psi}_{\mu }{\!}^{i}\,\gamma ^{\mu \nu }\psi _{\nu }{}^{j}+\varepsilon
^{ik}\,t_{k}{}^{j}\,\bar{\psi}_{\mu \,i}\,\gamma ^{\mu \nu }\psi _{\nu
\,j}\, \big]\,+6\,q^{2}\,e\,,
\end{align}
where $F(A)_{\mu \nu }=\partial _{\mu }A_{\nu }-\partial _{\nu }A_{\mu }$
and $e=\det (e_{\mu }{\!}^{a})$. The derivative of the gravitino fields is
covariant with respect to local Lorentz and electromagnetic gauge
transformations, and reads 
\begin{align}  \label{eq:covariant-graviton-der3}
\mathcal{D}_{\mu }\psi _{\nu }{\!}^{i}=& \;\big(\partial _{\mu }-\tfrac{1}{4}
\omega _{\mu }{\!}^{ab}\gamma _{ab}\big)\psi _{\nu }^{i}-\tfrac{1}{2}\sqrt{2}
\,q\,A_{\mu }\,t^{i}{\!}_{j}\,\psi _{\nu }{}^{j}\,,  \nonumber \\[1mm]
\mathcal{D}_{\mu }\psi _{\nu \,i}=& \;\big(\partial _{\mu }-\tfrac{1}{4}
\omega _{\mu }{\!}^{ab}\gamma _{ab}\big)\psi _{\nu \,i}-\tfrac{1}{2}\sqrt{2}
\,q\,A_{\mu }\,t_{i}{}^{j}\,\psi _{\nu \,j}\,,
\end{align}
where $\omega _{\mu }{\!}^{ab}$ is the spin connection whose definition will
be discussed momentarily. The matrices ${\textstyle\frac{1}{2}}\gamma_{ab} ={%
\textstyle\frac{1}{4}}[\gamma _{a},\gamma _{b}]$ are the Lorentz group
generators in the spinor representation. As mentioned already, $t^{i}{\!}%
_{j} $ is the anti-hermitian traceless generator of the electromagnetic
gauge transformations, which is an abelian subgroup of $\mathrm{SU}(2)$. It
is normalized to $t^{i}{\!}_{j}\,t_{i}{}^{j}=2$, where $t_{i}{}^{j}$ denotes
the complex conjugate of $t^{i}{\!}_{j}$. This implies the convenient
identities, 
\begin{equation}  \label{eq:t-properties}
t^{i}{\!}_{j}\,t_{k}{}^{j}=\delta ^{i}{\!}_{k}=-t^{i}{\!}_{j}\,t^{j}{\!}
_{k}\qquad \varepsilon _{ik}\,t^{k}{\!}_{j}=\varepsilon _{jk}\,t^{k}{\!}
_{i}\,,\qquad t_{i}{}^{j}\equiv (t^{i}{\!}_{j})^{\ast }=\varepsilon
_{ik}\,\varepsilon ^{jl}\,t^{k}{}_{l}\,.
\end{equation}
These identities do not lead to a unique choice for $t^{i}{\!}_{j}$; this is
consistent with the fact that the matrix can be redefined by applying a
uniform chiral $\mathrm{SU}(2)$ field redefinition on the spinors.

The spin connection $\omega _{\mu }{\!}^{ab}$ is derived from the
supercovariant torsion constraint, 
\begin{equation}
\mathcal{D}(\omega )_{\mu }\,e_{\nu }{\!}^{a}-\mathcal{D}(\omega )_{\nu
}\,e_{\mu }{\!}^{a}=\tfrac{1}{2}\big[\bar{\psi}_{\mu \,i}\gamma ^{a}\psi
_{\nu }{\!}^{i}+\bar{\psi}_{\mu }{\!}^{i}\gamma ^{a}\psi _{\nu \,i}\big]\,,
\label{eq:spin-connection}
\end{equation}%
where the Lorentz covariant derivative reads $\mathcal{D}(\omega)_{\mu}
\,e_{\nu }{\!}^{a}=\partial _{\mu }e_{\nu }{\!}^{a}-\omega
_{\!}^{ab}\,e_{\nu \,b}$. This constraint can be solved algebraically and
leads to, 
\begin{equation}
\omega _{\mu }{\!}^{ab}={\textstyle\frac{1}{2}}e_{\mu }{}^{c}\big(\Omega
^{ab}{}_{c}-\Omega ^{b}{}_{c}{}^{a}-\Omega _{c}{}^{ab}\big)\,,
\label{eq:sol-spin-connection}
\end{equation}
where the $\Omega _{ab}{}^{c}$ are the \textit{objects of anholonomity}. The
affine connection equals $\Gamma _{\mu \nu }{}^{\rho }=e_{a}{\!}^{\rho }\, 
\mathcal{D}_{\mu }(\omega )\,e_{\nu }{\!}^{a}$, and ensures the validity of
the vielbein postulate. In the absence of torsion, where the right-hand side
of \eqref{eq:spin-connection} vanishes, we have 
\begin{equation}
\Omega _{ab}{}^{c}=e_{a}{\!}^{\mu }\,e_{b}{\!}^{\nu }\,(\partial _{\mu
}e_{\nu }{\!}^{c}-\partial _{\mu }e_{\nu }{\!}^{c})\,.
\end{equation}
The corresponding expression for the affine connection is then equal to the
Christoffel connection.

The curvature associated with the spin connection equals 
\begin{equation}
R_{\mu \nu }{\!}^{ab}(\omega )=\partial _{\mu } \omega _{\nu }{\!}%
^{ab}-\partial _{\nu }\omega _{\mu }{\!}^{ab}-\omega _{\mu }{\!}
^{ac}\,\omega _{\nu \,c}{}^{b}+\omega _{\nu }{\!}^{ac}\,\omega _{\mu
\,c}{}^{b}{}\,,  \label{eq:Riemann-curv}
\end{equation}
which satisfies the Bianchi identity $\mathcal{D}(\omega )_{[\mu }\,R_{\nu
\rho ]}{\!}^{ab}(\omega )=0$. After converting the tangent-space indices in $%
R_{\mu \nu }{\!}^{ab}(\omega )$ to world indices, it will be equal to the
Riemann tensor, up to terms quadratic in the gravitino fields that originate
from the right-hand side in \eqref{eq:spin-connection}. Its contractions, 
\begin{equation}  \label{eq:Ricci}
R_{\mu }^{\,a}(e,\omega )=e_{b}{\!}^{\nu }\,R_{\mu \nu }{\!}^{ab}(\omega
)\,,\qquad R(e,\omega )=e_{a}{\!}^{\mu }\,e_{b}{\!}^{\nu }\,R_{\mu \nu }{\!}%
^{ab}(\omega )\,,
\end{equation}
yield the Ricci tensor and scalar, up to gravitino terms. Substituting the
solution of \eqref{eq:spin-connection} into $R(e,\omega )$ yields the Ricci
scalar up to terms \textit{quartic} in the gravitino fields.

Let us now list the supersymmetry transformation rules, 
\begin{align}  \label{eq:susy-transformations}
\delta e_\mu{\!}^a =\,& \bar\epsilon^i \gamma^a \psi_{\mu i}+ \bar\epsilon_i
\gamma^a \psi_\mu{}^i\,,  \nonumber \\[1mm]
\delta \psi_\mu{\!}^i =\;& 2\, \mathcal{D}_\mu \epsilon^i
-\tfrac14F(A)_{\rho\sigma} \gamma^{\rho\sigma} \gamma_\mu
\,\varepsilon^{ij}\, \epsilon_j + \sqrt{2} \,q \,\varepsilon^{ij} \,t_j{}^k
\,\gamma_\mu \epsilon_k\,,  \nonumber \\[1mm]
\delta \psi_{\mu\,i} =\;& 2\, \mathcal{D}_\mu \epsilon_i
-\tfrac14F(A)_{\rho\sigma} \gamma^{\rho\sigma} \gamma_\mu
\,\varepsilon_{ij}\, \epsilon^j + \sqrt{2}\, q \,\varepsilon_{ij}\, t^j{}_k
\,\gamma_\mu \epsilon^k\,,  \nonumber \\[1mm]
\delta A_\mu =\;& 2\,\big( \varepsilon^{ij}\, \bar\epsilon_i\psi_{\mu j} +
\varepsilon_{ij}\, \bar\epsilon^i\,\psi_\mu{}^j \big)\,,
\end{align}
where in the gravitino transformations we suppressed terms cubic in the
gravitino fields. The covariant derivatives of the supersymmetry parameters
are given by 
\begin{align}  \label{eq:covariant-epsilon-der}
\mathcal{D}_\mu \epsilon^i =&\; \big( \partial_\mu - \tfrac14 \omega_\mu{\!}%
^{ab} \gamma_{ab}\big) \epsilon^i - \tfrac12\sqrt{2}\, q\, A_\mu \,t^i{\!}_j
\,\epsilon^j \,,  \nonumber \\[1mm]
\mathcal{D}_\mu\epsilon_i =&\; \big( \partial_\mu - \tfrac14 \omega_\mu{\!}
^{ab} \gamma_{ab}\big) \epsilon_i - \tfrac12\sqrt{2}\, q\, A_\mu \,t_i{}^j
\,\epsilon_j \,.
\end{align}

The Lagrangian \eqref{eq:Lagrangian} is invariant under space-time
diffeomorphisms, supersymmetry, local Lorentz transformations, and
electromagnetic gauge transformations, whose infinitesimal transformations
will close under commutation. Of particular interest is the commutator of
two supersymmetry transformations, which closes into the diffeomorphism with
parameter $\xi ^{\mu }$, the local Lorentz transformations and the
electromagnetic gauge transformations, but only modulo the gravitino field
equations, 
\begin{equation}  \label{eq:QQ-commutator}
\lbrack \delta (\epsilon _{1}),\delta (\epsilon _{2})]=\xi ^{\mu }{\hat{D}}%
_{\mu }+\delta _{\mathrm{L}}(\varepsilon )+\delta (\Lambda )\,,
\end{equation}
where the derivative is fully covariant with respect to all the symmetries.
This implies that there is a contribution from $\xi ^{\mu }$ times each of
the connections that contribute. The explicit variations are simply
additional and they do not involve the connections. The parameters of the
various infinitesimal transformations on the right-hand side are given by 
\begin{align}  \label{eq:parameters-susy-alg}
\xi ^{\mu }=& \,2\,\bar{\epsilon}_{2}{}^{i}\gamma ^{\mu} \epsilon_{1i} + 
\text{h.c.}\,,  \nonumber \\
\varepsilon ^{ab}=& \, \varepsilon _{ij}\, \bar{\epsilon}_{1}{}^{i}
\epsilon_{2}{}^{j}\,F^{ab +}+\text{h.c.}\,,  \nonumber \\
\Lambda =& \, 4\, \varepsilon_{ij}\, \bar\epsilon_2{}^i\, \epsilon_1{}^j + 
\text{h.c.}\,,
\end{align}%
where the first term proportional to $\xi ^{\mu }$ denotes a supercovariant
translation, i.e. a general coordinate transformation with parameter $\xi
^{\mu }$, suitably combined with field-dependent gauge transformations so
that the result is supercovariant.

We will be interested in solutions that have full or partial supersymmetry.
The fully supersymmetric solution is well known and we will briefly refer to
it at the end of this section. There exist many solutions with partial
supersymmetry. Well-known examples are, for instance, the extremal
Reissner-Nordstr\"{o}m black holes solutions, which are invariant under half
the supersymmetries \cite{Romans:1991nq}. However, the main objective of
this paper is to analyze the possible supersymmetry of wormhole solutions
belonging to the class constructed in \cite{Anabalon:2018rzq}.

When a bosonic field configuration is fully or partially supersymmetric, it
implies that all or some of the supersymmetry transformations of the
fermions are vanishing. The transformations that vanish are characterized by
certain spinorial parameters that are known as generalized Killing spinors.
The only spinors that we are dealing with in this particular case are the
gravitini, so we have to simply analyse their supersymmetry transformation,
which amounts to deriving possible solutions for $\epsilon^{i}$ and $%
\epsilon_{i}$ of the equations, 
\begin{align}  \label{eq:killing-spinor-eqs}
2\,\mathcal{D}_{\mu }\epsilon ^{i}-\tfrac{1}{4}F(A)_{\rho \sigma }\gamma
^{\rho \sigma }\gamma _{\mu }\,\varepsilon ^{ij}\,\epsilon _{j}+\sqrt{2}
\,q\,\varepsilon ^{ij}t_{j}{}^{k}\,\gamma _{\mu }\epsilon _{k}=& \;0\,, 
\nonumber \\[1mm]
2\,\mathcal{D}_{\mu }\epsilon _{i}-\tfrac{1}{4}F(A)_{\rho \sigma }\gamma
^{\rho \sigma }\gamma _{\mu }\,\varepsilon _{ij}\,\epsilon ^{j}+\sqrt{2}
\,q\,\varepsilon _{ij}t^{j}{}_{k}\,\gamma _{\mu }\epsilon ^{k}=& \;0\,.
\end{align}
It is convenient to first consider an integrability condition for these
differential equations, which follows by applying a second derivative $%
\mathcal{D}_{\nu }$ and anti-symmetrizing over the indices $\mu$ and $\nu$.
The resulting equations take the following form, 
\begin{align}  \label{eq:susy-integrability}
\Xi _{\mu \nu }{\!}^{i}\equiv\,& \mathcal{D}_{\mu }\delta\psi_{\nu}{\!}^{i} -%
\mathcal{D}_{\nu }\delta \psi_{\mu }{\!}^{i} = \big[R(\omega )_{\mu \nu }{\!}%
^{ab}\,\gamma _{ab}-4\,q^{2}\,\gamma _{\mu \nu }+\tfrac{1}{8}F_{\rho \sigma
}\,F_{\lambda \tau }\,\gamma ^{\rho \sigma }\,\gamma _{\lbrack \mu }\,\gamma
^{\lambda \tau }\gamma _{\nu ]}\, \big]\epsilon ^{i}  \nonumber \\[1mm]
& +\tfrac{1}{2}\sqrt{2}\,q\big[4\,F_{\mu \nu }-F_{\rho \sigma }\,\gamma
^{\rho \sigma }\,\gamma _{\mu \nu }-\gamma _{\lbrack \mu }\,F_{\rho \sigma
}\,\gamma ^{\rho \sigma }\gamma _{\nu ]}\,\big]t^{i}{\!}_{j}\,\epsilon
^{j}+(\nabla _{\lbrack \mu }F_{\rho \sigma })\,\gamma ^{\rho \sigma }\gamma
_{\nu ]}\,\varepsilon ^{ij}\epsilon _{j}=0\,,  \nonumber \\[2mm]
\Xi _{\mu \nu i}\equiv\,& \mathcal{D}_{\mu }\delta \psi_{\nu i} -\mathcal{D}%
_{\nu } \delta \psi _{\mu i} = \big[R(\omega )_{\mu \nu }{\!}^{ab}\, \gamma
_{ab}-4\,q^{2}\,\gamma _{\mu \nu }+\tfrac{1}{8}F_{\rho \sigma }\,F_{\lambda
\tau }\,\gamma ^{\rho \sigma }\,\gamma _{\lbrack \mu }\,\gamma ^{\lambda
\tau }\gamma _{\nu ]}\,\big] \epsilon _{i}  \nonumber \\
& +\tfrac{1}{2}\sqrt{2}\,q\big[4\,F_{\mu \nu }-F_{\rho \sigma }\,\gamma
^{\rho \sigma }\,\gamma _{\mu \nu }-\gamma _{\lbrack \mu }\,F_{\rho \sigma
}\,\gamma ^{\rho \sigma }\gamma _{\nu ]}\,\big]t_{i}{}^{j}\,\epsilon
_{j}+(\nabla _{\lbrack \mu }F_{\rho \sigma })\,\gamma ^{\rho \sigma }\gamma
_{\nu ]}\,\varepsilon _{ij}\epsilon ^{j}=0\,,
\end{align}
where the covariant derivative $\nabla _{\mu }$ contains only the
Christoffel connection.

To analyze the above equations it is convenient to switch from two Majorana
spinors to a single Dirac spinor. To do so, one first chooses, without loss
of generality, the charge matrix $t^i{}_j$ to be equal to $\mathrm{diag}(%
\mathrm{i},-\mathrm{i})$. Subsequently one defines 
\begin{equation}  \label{eq:complex-susy}
\chi \equiv \epsilon^1 + \epsilon_2\,,\qquad \Xi _{\mu \nu}\chi \equiv \Xi_
{\mu\nu}{\!}^1 + \Xi _{\mu \nu\,2}\,.
\end{equation}
Now $\chi$ is no longer a Majorana spinor, because under charge conjugations
it will lead to another independent spinor $\epsilon_1+\epsilon^2$. Since
the two spinors are related by charge conjugation, it suffices to only
consider the quantities $\Xi _{\mu \nu}$, which constitute six different $%
4\times4$ matrices acting on the 4-component Dirac spinor $\chi$, defined by %
\eqref{eq:complex-susy}.

With these redefinitions the Killing spinor equations %
\eqref{eq:killing-spinor-eqs} and the integrability condition %
\eqref{eq:susy-integrability} reads as follows, 
\begin{align}  \label{eq:complex-killing-spinor-eqs}
&2\,\mathcal{D}_{\mu }\chi+ \tfrac{1}{4}F(A)_{\rho\sigma }
\gamma^{\rho\sigma} \gamma _{\mu }\gamma^5 \,\chi-\sqrt{2} \,\mathrm{i}
q\,\gamma _{\mu} \gamma^5\chi =0\,, \\[3mm]
&\Xi _{\mu \nu}\chi = \big[R(\omega )_{\mu \nu }{\!}^{ab}\,\gamma
_{ab}-4\,q^{2}\,\gamma _{\mu \nu }+\tfrac{1}{8}F_{\rho \sigma }\,F_{\lambda
\tau }\,\gamma ^{\rho \sigma }\,\gamma _{\lbrack \mu }\,\gamma ^{\lambda
\tau }\gamma _{\nu ]}\, \big]\chi  \nonumber \\[1mm]
& \qquad +\tfrac{1}{2}\sqrt{2}\,\,\mathrm{i} q\big[4\,F_{\mu \nu } -F_{\rho
\sigma }\,\gamma^{\rho \sigma }\, \gamma _{\mu \nu }-\gamma _{\lbrack \mu
}\, F_{\rho \sigma}\,\gamma ^{\rho \sigma }\gamma _{\nu ]}\,\big] \,\chi
-(\nabla _{\lbrack \mu }F_{\rho \sigma })\,\gamma ^{\rho \sigma }\gamma
_{\nu ]} \gamma^5\,\chi =0\,,  \nonumber
\end{align}
The covariant derivative of $\chi$ follows from %
\eqref{eq:covariant-epsilon-der}, 
\begin{equation}  \label{eq:cov-der-chi}
\mathcal{D}_\mu \chi = \big( \partial_\mu - \tfrac14 \omega_\mu{\!}^{ab}
\gamma_{ab} - \tfrac12\sqrt{2}\, \mathrm{i} q\, A_\mu \big) \chi \,.
\end{equation}
We recall that all fermionic fields have been suppressed on the right-hand
side of the equations \eqref{eq:killing-spinor-eqs} and %
\eqref{eq:cov-der-chi}, because we will be dealing with purely bosonic
backgrounds when exploring the possible supersymmetry of wormhole solutions.

The maximally supersymmetric solution has vanishing $A_{\mu}$, so that the
integrability relation then takes the form $R(\omega)_{\mu \nu }{\!}%
^{ab}=4\,q^{2}\,e_{\mu }{\!}^{[a}\,e_{\nu }{\!}^{b]}$. This equation implies
that the supersymmetric field configuration is just an anti-de Sitter
space-time with AdS radius $\ell$ given by 
\begin{equation}  \label{eq:ADSr}
\ell^{-1} = \sqrt2\,\vert q\vert\,.
\end{equation}
In the following sections we will consider a class of wormhole solutions
that can be partially supersymmetric. Their possible supersymmetry will be
investigated by analyzing the equations \eqref{eq:complex-killing-spinor-eqs}%
.


\section{Maxwell-Einstein-AdS wormholes}

\label{sec:einst-maxw-wormhole} \setcounter{equation}{0} 
Following \cite{Anabalon:2018rzq}, we consider a class of four-dimensional
space-time metrics expressed into two different functions, $f(r)$ and $h(r)$%
, 
\begin{equation}  \label{eq:class-of-metrics}
ds^{2}=\frac{4\,\ell ^{4}\,dr^{2}}{\sigma ^{2}f(r)}+h(r)\big[-\cosh
^{2}\theta \,dt^{2}+d\theta ^{2}\big]+f(r)(du+\sinh \theta dt)^{2}\,,
\end{equation}
where $\ell$ denotes the AdS radius. When considering supersymmetry we will
also need a corresponding set of vierbeine, for which we make the following
choice, 
\begin{align}  \label{eq:vierbeine}
e^{0}& =\sqrt{h(r)}\,\cosh \theta \,dt\,,  \nonumber \\[.5mm]
e^{1}& =\frac{1}{\sigma \,q^{2}\sqrt{f(r)}}\,dr\,,  \nonumber \\[.3mm]
e^{2}& =\sqrt{h(r)}\,\,d\theta \,,  \nonumber \\[2.1mm]
e^{3}& =\sqrt{f(r)}\,\big(du+\sinh \theta \,dt\big)\,.
\end{align}
For 
\begin{equation}
\sigma =4\,,\quad f(r)=h(r)=\tfrac{1}{4}\ell ^{2}(r^{2}+1)\, ,
\end{equation}
this defines a global $\mathrm{AdS}_{4}$ space-time. Its topology is trivial
because the coordinates cover the full ${\mathbb{R}}^{4}$.

However, it is possible to impose identifications on surfaces that are
orthogonal to $\partial_{r}$ so that one obtains a constant curvature
wormhole. In this case the space-time is only locally $\mathrm{AdS}_{4}$ and
has two conformal boundaries located at $r=\pm \infty $. The relevant
identification in the Lorentzian case is $u\sim u+a$, which for constant $r$
yields the three-dimensional Cousaert-Henneaux space-time \cite%
{Coussaert:1994tu}. This identification obviously introduces a
non-contractible cycle in space-time. In the case at hand, the location of
the throat is at $r=0$, when the non-contractible circle has minimal
(geodesic) length. The perimeter of the throat given by $a$ is an extra
parameter of the metric that is encoded in the range of the compact
coordinate $u\in \lbrack 0,a]$. This is not sufficient to prove that this
geometry defines a traversable wormhole. For that one needs to send
information from one side of the throat to the other, and one has to check
that there are no closed time-like curves. The later was addressed in \cite%
{Anabalon:2018rzq} for the uncharged case and the same argument applies
here; the former will be studied in section 5 below.

Since this field configuration is a solution of the Einstein-Maxwell system
with a cosmological term, it can also be a solution of pure $N=2$
supergravity, which means that it is a solution of its bosonic field
equations that follow from the Lagrangian \eqref{eq:Lagrangian}. These
combined field equations that it satisfies will therefore take the form 
\begin{align}  \label{eq:field-equations}
& \partial _{\mu }\big(e\,F^{\mu \nu }\big)=0\,\,,  \nonumber \\
& R_{\mu \nu }-\tfrac{1}{2}g_{\mu \nu }R+\tfrac{1}{2}\big[F_{\mu\rho}\,
F_{\nu }{}^{\rho }-\tfrac{1}{4}g_{\mu \nu }F_{\rho \sigma }F^{\rho\sigma} %
\big]+6\,q^{2}\,g_{\mu \nu }=0\,,
\end{align}
where $\ell^{-1}= \sqrt{2}\,\vert q\vert$ and $F_{\mu\nu} =0$.

Let us now move to a more complicated metric where the functions $f(r)$ and $%
h(r)$ are equal to 
\begin{equation}  \label{eq:wormhole-f-h}
f(r)=\,\frac{2}{q^{2}\,\sigma^{2}}\,\frac{r^{4}+( 6-\sigma)
r^{2}+m\,r+\sigma -3}{r^{2}+1}-\frac{Q^{2}+P^{2}}{r^{2}+1}\,,\qquad h(r)= 
\frac{1}{2\,q^{2}\,\sigma }\left( r^{2}+1\right) \,,
\end{equation}
and construct a corresponding solution of the above equations. Here $Q$ and $%
P$ are electric and magnetic charge \textit{parameters} that will determine
the physical charges (whose definition requires to properly account for
wormhole topology) and the corresponding electric and magnetic fields of the
solution. These charges are induced because the second field equation %
\eqref{eq:field-equations} requires the presence of electric and magnetic
fields, which will be given momentarily. Note, however, that we still retain
the homogeneous Maxwell equations, because the only charged sources are the
gravitini, which are not included in the bosonic background solution. In
addition the metric depends on two integration constants denoted by $m$ and $%
\sigma $. The parameter $m$ is proportional to the mass of the space-time,
while $\sigma$ is related to the warping of the asymptotic region. More
details can be found in \cite{Anabalon:2018rzq}.

The solution of \eqref{eq:field-equations} for the vector potential is given
by 
\begin{equation}
A=\Phi \left( r\right) \,\big(du+\sinh \theta \,dt\big)\text{ }\,,
\label{eq:vector-field}
\end{equation}
with $\Phi (r)$ equal to 
\begin{equation}
\Phi \left( r\right) =\frac{2\,Q\,r+P(1-r^{2})}{r^{2}+1}\,.
\label{eq:eq:Phis}
\end{equation}
It turns out that \eqref{eq:vector-field} is invariant under the isometries
given below in \eqref{eq:killing-vectors}. Obviously the vector potential $%
A_{\mu }$ describes an electric and a magnetic field component. Its field
strength in the adopted coordinate system is equal to 
\begin{align}
F_{ru}=& \,\frac{2(1-r^{2})Q-4\,r\,P}{(r^{2}+1)^{2}}\,,  \nonumber
\label{eq:field-strengths} \\
F_{rt}=& \,\frac{2(1-r^{2})Q-4\,r\,P}{(r^{2}+1)^{2}}\,\sinh \theta \,, 
\nonumber \\
F_{\theta t}=& \,\frac{2\,Q\,r+P(1-r^{2})}{r^{2}+1}\,\cosh \theta \,.
\end{align}
The possible existence of a non-contractible cycle requires that $f(r)$ must
be positive everywhere.\footnote{
A straightforward analysis shows that $f(r)$ never vanishes provided 
\begin{align}  \label{eq:regularity}
& X=3(Q^{2}+P^{2})\ell ^{-2}\leq 1\,, \qquad \frac{12+12\sqrt{1-X}}{1+X+%
\sqrt{1-X}}>\sigma > \frac{12-6\sqrt{1-X}}{1+X+\sqrt{1-X}}\,, \\[1mm]
& \vert m \vert <\frac{\sqrt{2}}{3\sqrt{3}}\,\frac{\sigma \left( 6-\sigma
\right) \sqrt{1-X}+24\sigma -\sigma ^{2}\left( 1+X\right) -72}{\sqrt{\sigma
\left( 1+\sqrt{1-X}\right) -6}}\,.  \nonumber
\end{align}
For these ranges of the parameters, the metric functions are everywhere
positive and regular and a non-trivial wormhole space-time will exist.} 
Asymptotically, for $r=\pm\infty$ the space-time is locally $\mathrm{AdS}_4$
with the following fall-off for the curvature tensor, 
\begin{equation}
R(\omega )_{\mu \nu }{\!}^{ab}=\left[ 2\,\ell ^{-2}+\mathcal{O}(r^{-2})%
\right] e_{\mu }{\!}^{[a}\,e_{\nu }{\!}^{b]}\,.
\end{equation}

The bosonic field configuration associated with global $\mathrm{AdS}_4$ is
invariant under the isometry group $\mathrm{SO}(3,2)$. This group is broken
for the deformed functions $f(r)$ and $h(r)$ specified in %
\eqref{eq:wormhole-f-h} and the electromagnetic fields %
\eqref{eq:field-strengths} to a subgroup generated by the following four
Killing vectors, 
\begin{align}  \label{eq:killing-vectors}
{\xi }_{[1]}=\,& \,\partial _{t}\,,  \nonumber \\
{\xi }_{[2]}=\,& \,\sin t\,\partial _{\theta }+\tanh \theta \,\cos t\,
\partial_{t}+\frac{\cos t}{\cosh \theta }\,\partial_{u}\,,  \nonumber \\
{\xi }_{[3]}=\,& \,\cos t\,\partial _{\theta }-\tanh \theta \,\sin
t\,\partial _{t}-\frac{\sin t}{\cosh\theta }\,\partial_{u}\,,  \nonumber \\
{\xi }_{[4]}=\,& \,\partial_{u}\,.
\end{align}
We note that when $u$ is compact the $SO(3,2)$ isometries of global $AdS_4$
are also broken to these four Killing vectors. The first three Killing
vectors generate the group $\mathrm{SO}(2,1)$, while the fourth isometry is
abelian and commutes with the first three. Not surprisingly, the two
functions given in \eqref{eq:wormhole-f-h} depend only on $r$ and are
therefore invariant under the four isometries. Our solution can be seen as a
deformation of $\mathrm{AdS}_{3}$ embedded in a four-dimensional space. The
deformation by the function $f(r)$ breaks the $\mathrm{SO}(2,2)\cong\mathrm{%
SO}(2,1)\times \mathrm{SO}(2,1)$ isometries to its subgroup $\mathrm{SO}%
(2,1)\times \mathrm{SO}(1,1)$.

We also calculate $\mathcal{L}_{\xi}\,e_{\mu }{\!}^{a}= \xi^{\nu} \partial
_{\nu}\,e_{\mu }{\!}^{a}+\partial _{\mu }\xi ^{\nu }\,e_{\nu}{\!} ^{a}$ for
each of the Killing vectors. As it turns out $\mathcal{L}_{\xi}\,e_{\mu}^{a}$
vanishes on all the vierbeine for the $\xi _{\lbrack 1]}$ and $\xi _{\lbrack
4]}$, while the non-trivial action of the other Killing vectors on the
vierbeine yields 
\begin{equation}
\begin{array}{rcl}
\mathcal{L}_{\xi _{[2]}}\,e^{0} \!\! & = & \!\!\frac{\displaystyle\cos t} {%
\displaystyle\cosh \theta } \,e^{2}\,, \\[2mm] 
\mathcal{L}_{\xi _{\lbrack 2]}}\,e^{2}\!\! & = & \!\!\frac{\displaystyle %
\cos t}{\displaystyle\cosh \theta} \,e^{0}\,,%
\end{array}
\qquad 
\begin{array}{rcl}
\mathcal{L}_{\xi _{\lbrack 3]}}\,e^{0} \!\! & = & \!\! - \frac{\displaystyle%
\sin t}{\displaystyle\cosh \theta }\, e^{2} \,, \\[2mm] 
\mathcal{L}_{\xi _{\lbrack 3]}}\,e^{2} \!\! & = & \!\!- \frac{\displaystyle%
\sin t}{\displaystyle\cosh \theta }\,e^{0} \,.%
\end{array}%
\end{equation}
Hence the vierbeine are not invariant under the diffeomorphisms generated by
the Killing vectors, but they are invariant under these diffeomorphisms when
accompanied by tangent-space transformations that are opposite to the ones
indicated above. On spinors these tangent transformations will take the form 
\begin{align}  \label{eq:tangent-tr}
\delta_{[2]} \psi = - \frac{\cos t} {2\,\cosh \theta} \,\gamma^0\gamma^2\,%
\psi \,,\qquad \delta_{[3]} \psi =\frac{\sin t} {2\,\cosh \theta}
\,\gamma^0\gamma^2\,\psi \,.
\end{align}
We will return to these compensating tangent-space transformation at the end
of sections \ref{sec:susy-wormhole-solutions}, where we will discuss the
corresponding invariances of the Killing spinors. Note that the
transformations $\xi_{[1]}$ and $\xi_{[4]}$ do not involve any compensating
tangent-space transformations.


\section{Supersymmetric wormholes}

\label{sec:susy-wormhole-solutions} \setcounter{equation}{0} 
To investigate whether the wormhole solutions can be supersymmetric, one may
first consider the integrability for the complex Killing spinors $\chi$,
which was presented in \eqref{eq:complex-killing-spinor-eqs}. In the actual
calculations we use the following representation for the gamma matrices, 
\begin{equation}  \label{eq:gamma-matrices}
\gamma ^{0}=-\mathrm{i} \begin{pmatrix} 0\! & \sigma _{2} \\ \sigma _{2} & 0
\end{pmatrix} \;\quad \gamma ^{1}=- \begin{pmatrix} \!\sigma _{3}\! & 0 \\ 0
& \sigma _{3} \end{pmatrix} \,,\quad \gamma ^{2}=\mathrm{i}\begin{pmatrix} 0
& -\sigma _{2} \\ \sigma _{2} & 0 \end{pmatrix} \,,\quad \gamma ^{3}=%
\begin{pmatrix} \sigma _{1} & 0 \\ 0 & \sigma _{1} \end{pmatrix} \,,
\end{equation}
where we remind the reader of the definition $\gamma^{5} = -\mathrm{i}%
\gamma^{0}\gamma^{1} \gamma^{2}\gamma^{3} = \mathrm{diag}( \sigma_2,
-\sigma_2)$.\footnote{
With these gamma matrices we can choose the charge conjugation matrix as $S=
S^{-1}=-S^\mathrm{T}$, so that the charge conjugate of a spinor $\psi$ is
equal to $S \,\bar\psi^\mathrm{\,T} = \psi^{*}$.}

A necessary condition for the existence of non-trivial Killing spinors is
that the determinant of each of the six $4\times4$ matrices $\Xi _{\mu\nu}$
defined in \eqref{eq:complex-susy} must vanish. As it turns out all six
determinants take the form of a constant times $(r^2+1)^{-6}$ times a
function $Z(r)$. This function also depends on the charges and the
integration constants $\sigma$ and $m$ in the metric based on %
\eqref{eq:wormhole-f-h} so the condition for supersymmetry is that $Z(r)$
must vanish. Explicit calculation shows that the function $Z(r)$ has the
following form, 
\begin{equation}  \label{eq:defZ}
Z(r)=Z_{2}\,r^{2}+Z_{1}\,r+Z_{0}\,,
\end{equation}
where $Z_2$, $Z_{1}$ and $Z_{0}$ are fairly complicated expressions that
contain the charges and integration constants. However, the integrability
condition should hold for any value of the radial coordinate $r$. Therefore
one concludes that $Z_2$, $Z_1$ and $Z_0$ should separately vanish. For $Z_2$
this leads to the equation, 
\begin{equation}  \label{eq:Z-2}
Z_{2}=\frac{\sigma ^{2}}{q^{4}}\left( mP+8Q-2Q\sigma \right)=0\,.
\end{equation}
Since the metric is singular when $\sigma$ vanishes, we conclude that 
\begin{equation}  \label{eq:mass}
m=\, \frac{2\,Q}{P}\,(\sigma -4)\,.
\end{equation}
When this equation is satisfied then $Z_1$ turns out to vanish identically.
Hence the only remaining condition follows from requiring that $Z_0$ must
vanish, 
\begin{equation}  \label{eq:Z-0}
Z_{0}= \frac{\big( P^{2}+Q^{2}\big)^2} {2\,P^{4\,}q^{6}}\, \big(%
2q^{2}\sigma^{2}P^{2}+( \sigma -4)^{2} \big)\, \big( -2 ( \sigma-4)
+P^{2}\sigma ^{2}q^{2} \big)^{2} =0\,,
\end{equation}
where we made again use of equation \eqref{eq:mass}. Combining the above
results one obtains the conditions 
\begin{equation}  \label{eq:susy-parameter-range}
P=\frac{1} {\vert q\vert \,\sigma }\,\sqrt{2\,(\sigma -4)}\,,\qquad m\,=
\vert q\vert\, \sigma \,Q\, \sqrt{2\,(\sigma -4)}\,.
\end{equation}
Supersymmetry thus implies $\sigma > 4$ which is the same result that was
found in \cite{Anabalon:2018rzq} by requiring holographic stability.%
\footnote{
It is possible to define $\sigma$ in terms of the charge parameters $Q$ and $%
P$, but there are two solutions: 
\[
\sigma_{\pm } =\frac{1}{q^2\,P^{2}} \left( 1\pm \sqrt{1-8 \,q^2 P^{2}}%
\right)\,. 
\]%
} 

Now that we have solved the integrability condition for the existence of
Killing spinors, let us proceed to an explicit determination of these
spinors. To appreciate the possible relevance of the identification $u\sim
u+a$ for supersymmetry, we determine the possible Killing spinors
explicitly. To solve the Killing spinor we use the Dirac spinor $\chi$
defined in \eqref{eq:complex-susy}. The Killing spinor equations for $\chi$
was already given in \eqref{eq:killing-spinor-eqs}. Substituting the
expression for the bosonic covariant derivative, it reads 
\begin{equation}  \label{eq:KSE}
\big[\partial _{\mu } -\tfrac{1}{4}\omega _{\mu }{\!}^{ab} \gamma_{ab}
-\tfrac12\sqrt{2}\,\mathrm{i} q A_{\mu } +\frac{1}{8}F_{\rho\sigma}\gamma^{%
\rho\sigma}\gamma _{\mu } \gamma ^{5} -\tfrac1{2}\sqrt{2} \mathrm{i} q
\gamma _{\mu}\gamma^5 \big] \chi =0\,.
\end{equation}
It is useful to first study the Killing spinors of global $\mathrm{AdS}_{4}$
in terms of the coordinates used throughout this paper and to observe the
effect of having a compact $u$-coordinate on supersymmetry. We suppress for
the moment the presence of $A_\mu$ and $F_{\mu\nu}$ in \eqref{eq:KSE}. In
this way we obtain the following four real Killing spinors, 
\begin{align}  \label{eq:AdS-Killing-spinors}
\chi _{1}{\!}^\mathrm{AdS}=&\, \begin{pmatrix} \sqrt{1+\sqrt{r^{2}+1}}\,
\big[\cosh\theta/2\, \cos t/2 -\sinh{\theta/2}\,\sin t/2 \big] \\[1mm]
-\sqrt{ -1+\sqrt{r^{2}+1}} \, \big[\cosh\theta/2\, \cos t/2
-\sinh{\theta/2}\,\sin t/2 \big] \\[1mm] \sqrt{1+\sqrt{r^{2}+1}} \,
\big[\cosh\theta/2\, \sin t/2 -\sinh{\theta/2}\,\cos t/2 \big] \\[1mm]
\sqrt{-1+\sqrt{r^{2}+1}}\, \big[\cosh\theta/2\, \sin t/2
-\sinh{\theta/2}\,\cos t/2 \big] \end{pmatrix}\,,  \nonumber \\[3mm]
\chi_{2}{\!}^\mathrm{AdS} =&\, \begin{pmatrix}
-\sqrt{1+\sqrt{r^{2}+1}}\,\big[\sinh \theta/2\, \cos t/2 +\cosh \theta/2\,
\sin t/2 \big] \\[1mm] \sqrt{-1+\sqrt{r^{2}+1}}\,\big[ \sinh\theta/2\, \cos
t/2 +\cosh \theta/2 \, \sin t/2 \big] \\[1mm]
\sqrt{1+\sqrt{r^{2}+1}}\,\big[\sinh\theta/2\, \sin {t/2} +\cosh \theta/2\,
\cos{t/2} \big] \\[1mm] \sqrt{-1+\sqrt{1+r^{2}}}\, \big[\sinh \theta/2\,\sin
{t/2} +\cosh \theta/2\,\cos {t/2} \big] \end{pmatrix} \,,  \nonumber \\[3mm]
\chi _{3}{\!}^\mathrm{AdS}=&\,e^{u/2} \, \begin{pmatrix}
\sqrt{-1+\sqrt{r^{2}+1}} \\[1mm] -\sqrt{1+\sqrt{r^{2}+1}} \\[1mm] 0 \\[1mm]
0 \end{pmatrix}\,, \qquad \chi _{4}{\!}^\mathrm{AdS}=e^{-u/2}\, %
\begin{pmatrix} 0 \\[1mm] 0 \\[1mm] \sqrt{-1+\sqrt{r^{2}+1}} \\[1mm]
\sqrt{1+\sqrt{r^{2}+1}} \end{pmatrix}\,.
\end{align}

However, we have to remember that we are constructing representations for 
\textit{complex} Killing spinors, so that the above spinors can be
multiplied by arbitrary complex normalization factors. Hence we are dealing
with eight independent Killing spinors, which will indeed provide a basis
for full $N=2$ supersymmetry, as is expected for a global $\mathrm{AdS}_{4}$
space-time.

A noteworthy feature in the context of the present paper is that the last
two Dirac spinors, $\chi_{3}^{\mathrm{AdS}}$ and $\chi _{4}^{\mathrm{AdS}}$,
are incompatible with a periodic coordinate $u$. Therefore, when dealing
with a non-contractible cycle $u\sim u+a$, half of the Killing spinors will
no longer be globally defined, so that this particular field configuration
must be regarded as a 1/2-BPS solution. At the same time, the equations of
motion will still be locally satisfied.

At this point one can invoke the supersymmetry algebra given by %
\eqref{eq:QQ-commutator}, which relates the commutator of two supersymmetry
transformations to the bosonic symmetries of the model. When choosing
supersymmetry parameters expressed in terms of linear combinations of the
Killing spinors $\epsilon^i$, one obtains all the bosonic transformations
that should be compatible with the supersymmetric background, and in
particular one would obtain the Killing vectors of $\mathrm{AdS}_{4}$.
However, when the Killing spinors are not all globally defined, then some of
the Killing vectors of the space-time will not be globally defined either.

Let us now continue and derive the Killing spinors for the non-constant
curvature wormhole with non-trivial electromagnetic fields. A lengthy
analysis shows that there exist only two Dirac, Killing spinors, so that the
number of Killing spinors is reduced to one half. Furthermore, these spinors
do no longer depend on the coordinate $u$, so that they are globally
defined. We will give the explicit expressions momentarily. It turns out
that the first and the second component of these spinors differ by an
overall function $G(r)$, whereas the third and the fourth component differ
by an overall function $\bar{G}(r)$ that equals the complex conjugate of $%
G(r)$. This function $G(r)$ is quite complicated and takes the following
form, 
\begin{equation}  \label{eq:xG-function}
G(r)= \frac{-1} {q^2\, \sigma\, h(r)} \, \frac{\big[\sqrt{f(r)} -2\sqrt{2}\,
q \,h(r)\big]\, \big[\sqrt{f(r)}- \mathrm{i}\Phi(r)\big]} {f ^\prime (r) +%
\mathrm{i}\sqrt{f(r)}\,\Phi^\prime (r) }\,.
\end{equation}
The two Dirac Killing spinors now take the form, 
\begin{align}  \label{eq:wormhole-killing-spinors}
\chi _1{\!}^\mathrm{WH}=&\,\alpha(r) \begin{pmatrix} e^{\mathrm{i}\beta
(r)}\big[\cosh\theta/2\, \cos t/2 -\sinh{\theta/2}\,\sin t/2 \big] \\[1mm]
-e^{\mathrm{i}\beta(r)}\,G(r) \big[\cosh\theta/2\, \cos t/2
-\sinh{\theta/2}\,\sin t/2 \big] \\[1mm]
e^{-\mathrm{i}\beta(r)}\big[\cosh\theta/2\, \sin t/2 -\sinh{\theta/2}\,\cos
t/2 \big] \\[1mm] e^{-\mathrm{i}\beta(r)}\,\bar{G}(r) \big[\cosh\theta/2\,
\sin t/2 -\sinh{\theta/2}\,\cos t/2 \big] \end{pmatrix} \,,  \nonumber \\%
[3mm]
\chi _2{\!}^\mathrm{WH}=&\,\alpha(r) \begin{pmatrix} -e^{\mathrm{i}\beta
(r)}\big[\sinh\theta/2\, \cos t/2 +\cosh{\theta/2}\,\sin t/2 \big] \\[1mm]
e^{\mathrm{i}\beta(r)}\,G(r)\big[\sinh\theta/2\, \cos t/2
+\cosh{\theta/2}\,\sin t/2 \big] \\[1mm] e^{-\mathrm{i}\beta
(r)}\big[\sinh\theta/2\, \sin t/2 +\cosh{\theta/2}\,\cos t/2 \big] \\[1mm]
e^{-\mathrm{i}\beta (r)}\,\bar{G}(r) \big[\sinh\theta/2\, \sin t/2
+\cosh{\theta/2}\,\cos t/2 \big] \end{pmatrix} \,,
\end{align}
where 
\begin{align}
\alpha(r)=&\, \frac{h^{1/4}(r)} {\sqrt{1+\vert G(r)\vert^2}}\,, \\
e^{2\mathrm{i}\beta(r)}=&\,\big(1+\tfrac1{2}\mathrm{i}\sqrt{\sigma -4}\, %
\big)\, \sqrt\frac{f(r)}{h(r)} \,\frac{1+\vert G(r)\vert^2}{1+G(r)^{2}}\,.
\end{align}
Therefore we find that there are two independent Dirac Killing spinors
(which in this case are actually complex). This solution is therefore
1/2-BPS. As before we can invoke the supersymmetry algebra, and verify that
one reproduces the Killing vectors \eqref{eq:killing-vectors}, which will be
globally defined. All this provides a non-trivial check of the correctness
of our results.

We can also determine how these Killing spinors transform under the
symmetries of the bosonic field configuration. As explained at the end of
section \ref{sec:susy-wormhole-solutions}, these symmetries take the form of
a linear combination of the isometries \eqref{eq:killing-vectors} and
certain tangent-space transformations that act on spinors according to %
\eqref{eq:tangent-tr}. The Killing spinors thus transform under both
transformations. As it turns out, the tangent space transformation will
cancel in this linear combination, and we are left with the following
transformations, 
\begin{align}  \label{eq:isometry-var-on-killing-spinors}
\delta_{[1]} \chi^\mathrm{WH} =&\, \tfrac12 \begin{pmatrix} 0& 1 \\[1mm] -1
& 0 \end{pmatrix} \,\chi^\mathrm{WH}\,,  \nonumber \\[1mm]
\delta_{[2]} \chi^\mathrm{WH} =&\, \tfrac12 \begin{pmatrix} -1 &0 \\[1mm] 0&
1 \end{pmatrix}\,\chi^\mathrm{WH} \,,  \nonumber \\[1mm]
\delta_{[3]} \chi^\mathrm{WH} =&\,- \tfrac12 \begin{pmatrix} 0& 1 \\[1mm] 1
& 0 \end{pmatrix}\chi^\mathrm{WH} \,,  \nonumber \\[1mm]
\delta_{[4]} \chi^\mathrm{WH} =&\, 0 \,.
\end{align}
where 
\begin{equation}  \label{eq:chi-column}
\chi^\mathrm{WH} = \begin{pmatrix} \chi_1{\!}^\mathrm{WH}\\[1mm]
\chi_2{\!}^\mathrm{WH} \end{pmatrix} \,.
\end{equation}
Obviously the Killing spinors thus transform according the two-dimensional
representation of $\mathrm{SO}(2,1)$.


\section{Geometric aspects of supersymmetric wormholes}

\label{sec:geometric-aspects} 
In the previous section we proved the existence of $1/2$-BPS wormhole
solutions in $N=2$ supergravity. Now we turn to a discussion of the
geometric properties of these space-times.

\begin{figure}[ht]
\centering
\includegraphics[scale=0.3]{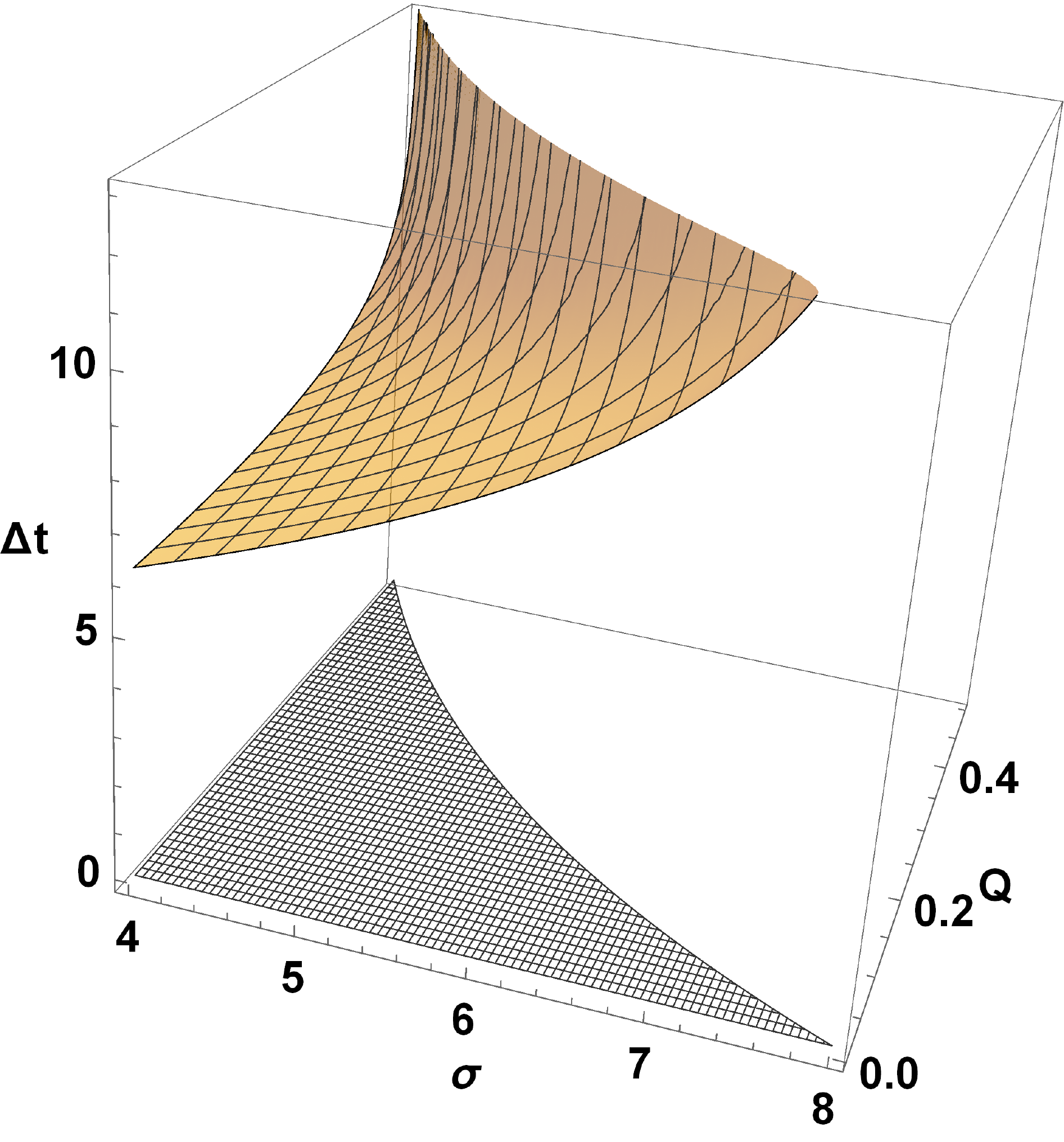}
\caption{Crossing time $\Delta{t}$ for a photon as a function of the charge $%
Q$ and the parameter $\protect\sigma$. The region where the metric is
regular and the wormhole is BPS corresponds to the shaded area in the lower
plane. This restriction originates from the bounds given in 
\eqref{eq:regularity} and the BPS conditions \eqref{eq:susy-parameter-range}%
. The crossing time remains finite, and starts growing as one approaches the
upper bound on $Q$.}
\label{fig:crossingBPS}
\end{figure}

The throat of the supersymmetric wormhole is located at the minimum of the
volume of the $t, r=\text{constant}$ surfaces. This is at the minimum of the
function $f(r)\cdot h(r)$. A plot with the time it takes for a photon to
cross the whole space-time, as seen by a geodesic observer located at $r= t
=\theta =0$ and constant $u$, is shown in Fig. \ref{fig:crossingBPS}. 

An important remark is now in order. The vector $\partial_t$, which is
asymptotically time-like for $\sigma \geq4$ may become space-like in the
interior of the wormhole when the following inequality holds, 
\begin{equation}
f\left( r\right) \sinh^{2}\theta-h\left( r\right) \cosh^{2}\theta >0\,.
\label{ineq}
\end{equation}
This would lead to an ergoregion, as happens in \cite{Lu:2008zs}, and tends
to be in contradiction with supersymmetry \cite{Townsend:2002yf}. However,
from the supersymmetry conditions (\ref{eq:susy-parameter-range}) one can
show in a straight-forwarded manner that the inequality (\ref{ineq}) cannot
be fulfilled, so that the asymptotically timelike Killing vector $%
\partial_{t}$ is actually timelike everywhere in the interior of the BPS
wormhole geometry.

The induced metric on the surfaces at constant $t$ and $\theta$ is equal to 
\begin{equation}
ds^{2}=\frac{dr^{2}}{q^{4}\sigma ^{2}f\left( r\right) }+f\left( r\right)
du^{2}=d\rho ^{2}+R^{2}\left( \rho \right) du^{2}\,,
\end{equation}
where the second equation is obtained by going to the proper radial
coordinate. The function $R(\rho)$ defines the radius of the circles
parameterized by the compact coordinate $u$. As $r\rightarrow \pm \infty $
one has $\rho\rightarrow \pm\infty$ and $R\left( \rho \right) \sim e^{\pm
\rho}$, as expected due to the locally $\mathrm{AdS}$ asymptotics. The
coordinate $\rho$ is such that $r=0$ implies $\rho =0$. Fig. \ref%
{fig:embsusy} shows the plot of the radial function $R(\rho )$ as a function
of $\rho $. The latter runs radially on the wormhole geometry and measures
the proper radial distance from the throat. 
\begin{figure}[th]
\centering
\includegraphics[scale=0.2]{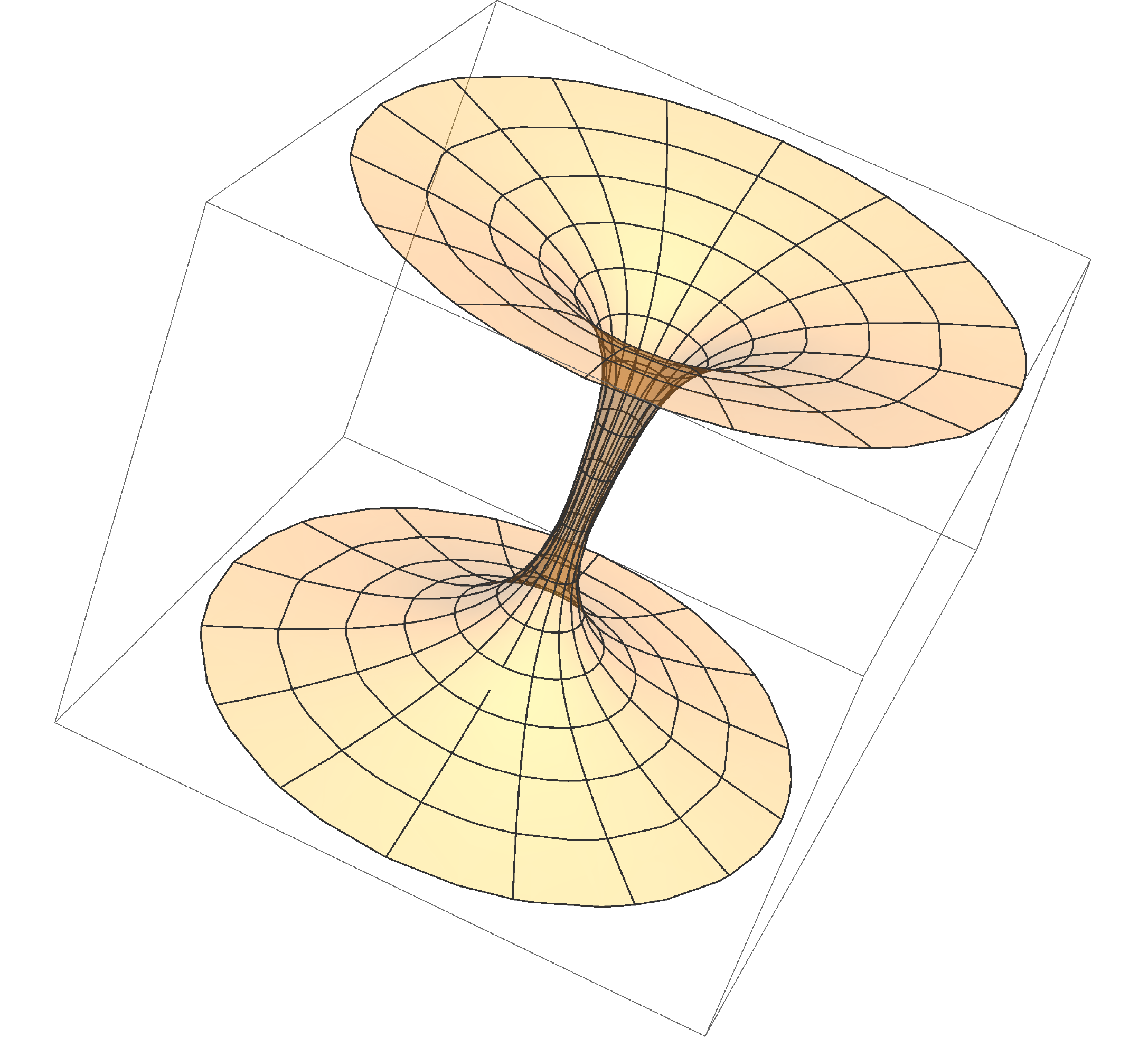}\qquad %
\includegraphics[scale=0.15]{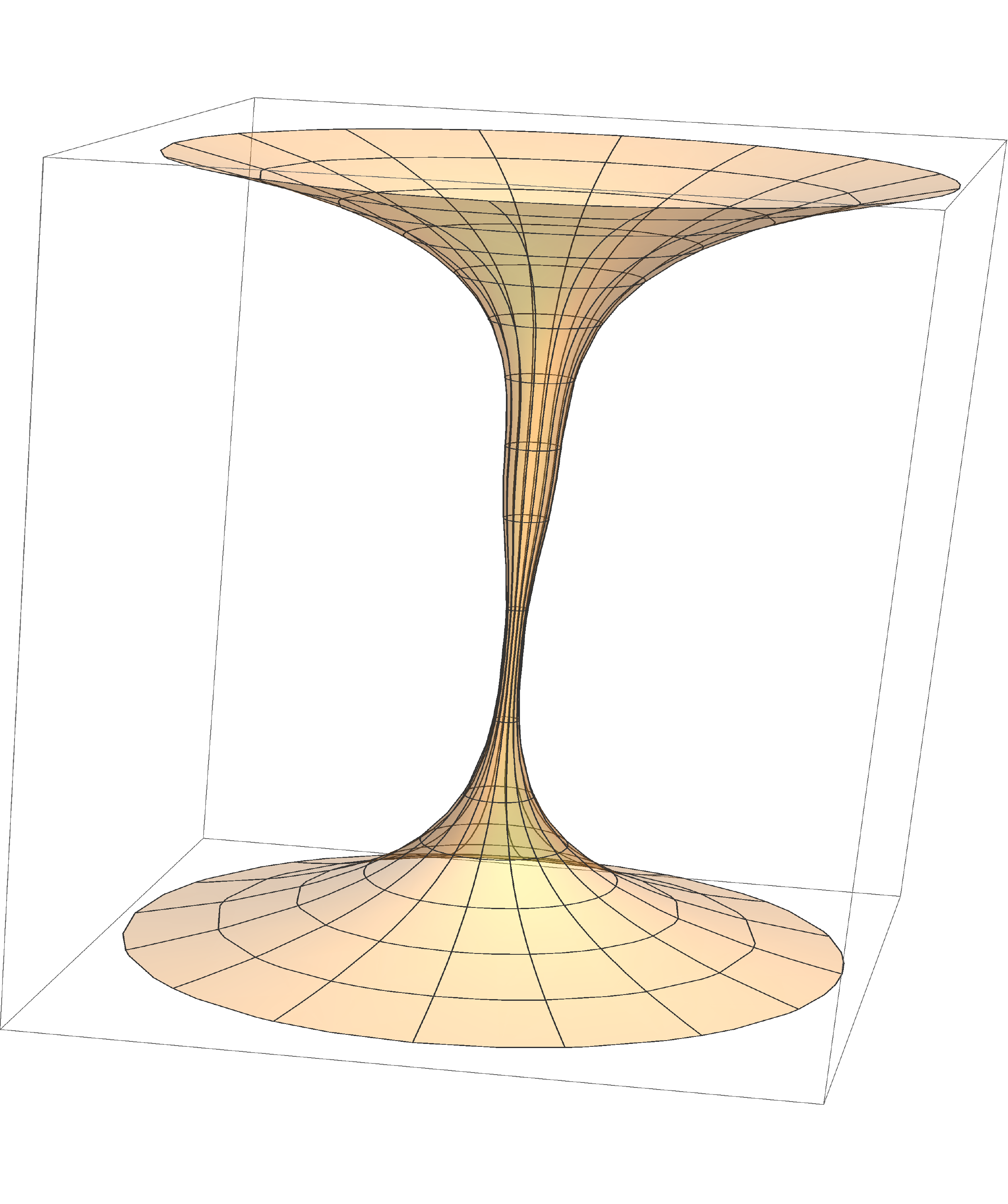}
\caption{Embedding of the charged supersymmetric wormhole with $Q=10^{-1}$
and $\protect\sigma =5$ (left panel) and $\protect\sigma =6$ (right panel).}
\label{fig:embsusy}
\end{figure}


\subsection{The geometry of the conformal boundary}

Let us pass to study the asymptotic region of the space-time. To this end,
it is enlightening to understand first pure $\mathrm{AdS}_4$ space-time. We
would like to clarify the relation between the global coordinate system we
use in this article and the standard global coordinate system where the
space-time is foliated by spheres. $\mathrm{AdS}$ is the Lorentzian
hyperboloid, where the AdS radius is set to one, $\ell =1$, 
\begin{equation}
-X_{0}{\!}^{2}-X_{1}^{2}+X_{2}{\!}^{2}+X_{3}{\!}^{2}+X_{4}{\!}^{2}=-1\,.
\end{equation}%
This constraint is solved by the global parametrization, 
\begin{eqnarray}
X_{0}^{S^{2}} &=&\sqrt{\rho ^{2}+1}\sin \tau \,, \qquad X_{1}^{S^{2}}=\sqrt{%
\rho ^{2}+1}\cos \tau \text{ },  \nonumber \\
X_{2}^{S^{2}} &=&\rho\,\sqrt{1-y^{2}}\cos \phi \text{ },\quad
X_{3}^{S^{2}}=\rho\, \sqrt{1-y^{2}}\sin \phi \text{ },  \nonumber \\
X_{4}^{S^{2}} &=&\rho \,y\,,
\end{eqnarray}
we shall call this the sphere foliation. It has the (universal covering)
metric 
\begin{equation}
ds_{S^{2}}^{2}=-\left( 1+\rho ^{2}\right) d\tau ^{2}+\frac{d\rho ^{2}}{%
1+\rho ^{2}}+\rho ^{2}\left( \frac{dy^{2}}{1-y^{2}}+\left( 1-y^{2}\right)
d\phi ^{2}\right) \text{ },
\end{equation}
where $y\in \lbrack -1,1]$, $\phi \in \lbrack 0,2\pi ]$, $\tau \in \lbrack
-\infty ,\infty ]$ and $\rho \in \lbrack 0,\infty ]$. Elsewhere in this
paper we have used another global parametrization, 
\begin{eqnarray}
X_{0}^{AdS_3} &=&\big[\sin t/2\, \sinh{u/2}\, \sinh{\theta/2} + \cos{t/2}%
\,\cosh{u/2}\,\cosh{\theta/2}\big] \sqrt{r^{2}+1} \,  \nonumber \\
X_{1}^{AdS_3} &=&\big[\sin t/2\,\cosh{u/2}\,\cosh{\theta/2}-\cos{t/2}\, \sinh%
{u/2}\,\sinh{\theta/2} \big] \sqrt{r^{2}+1}\,,  \nonumber \\
X_{2}^{AdS_3} &=&\big[\sin{t/2}\,\sinh{u/2}\,\cosh{\theta/2} -\cos{t/2}%
\,\cosh{u/2}\,\sinh{\theta/2} \big]\sqrt{r^{2}+1}\,,  \nonumber \\
X_{3}^{AdS_3} &=&\big[\sin{t/2}\,\cosh{u/2}\,\sinh{\theta/2} + \cos{t/2}%
\,\sinh{u/2}\,\cosh{\theta/2}\big]\sqrt{r^{2}+1}\,,  \nonumber \\
X_{4}^{AdS_3} &=&r\,.
\end{eqnarray}
The metric is now foliated by $AdS_{3}$ space-times, 
\begin{equation}
ds_{\mathrm{AdS}_{3}}^{2}=\frac{dr^{2}}{r^{2}+1}+\frac{\left( r^{2}+1\right) 
}{4}\left( -\cosh \left( \theta \right) ^{2}dt^{2}+d\theta ^{2}+\left(
du+\sinh \theta dt\right) ^{2}\right)  \label{AdS4}
\end{equation}%
where $r\in \lbrack -\infty ,\infty ]$, $\theta \in \lbrack -\infty ,\infty
] $, $t\in \lbrack -\infty ,\infty ]$ and global AdS has $u\in \lbrack
-\infty ,\infty ]$. The space-time is non-simply connected when $u$ is
periodic, $u\in \lbrack 0,a]$.

These two foliations are obviously related by a change of coordinates: 
\begin{eqnarray}
u &=&\frac{1}{2}\ln \left( \frac{1+2\rho ^{2}-\rho ^{2}y^{2}-2\rho \cos
\left( \tau -\phi \right) \sqrt{\left( 1+\rho ^{2}\right) \left(
1-y^{2}\right) }}{1+2\rho ^{2}-\rho ^{2}y^{2}+2\rho \cos \left( \tau -\phi
\right) \sqrt{\left( 1+\rho ^{2}\right) \left( 1-y^{2}\right) }}\right) 
\text{ },  \label{cc1} \\
\sinh \theta &=&\frac{2\rho \sin \left( \phi -\tau \right) \sqrt{\left(
1+\rho ^{2}\right) \left( 1-y^{2}\right) }}{1+\rho ^{2}y^{2}}\text{ }, \\
\sin t &=&\frac{\sin (2\tau )\left( 1+\rho ^{2}\right) -\sin (2\phi )\rho
^{2}\left( 1-y^{2}\right) }{\sqrt{\left( 1+2\rho ^{2}-\rho ^{2}y^{2}\right)
^{2}-4\rho ^{2}\left( 1-y^{2}\right) \left( 1+\rho ^{2}\right) \cos \left(
\tau -\phi \right) ^{2}}}\text{ }, \\
r &=&\rho y\text{ }.  \label{cc4}
\end{eqnarray}%
It follows that positive $r$ corresponds to the northern hemisphere of the $%
\mathbb{S}^{2}$ and negative $r$ to the southern hemisphere in the sphere
foliation. Therefore the boundary at $r>0$ ($y>0$) should be in some form
connected with the boundary at $r<0$ ($y<0$) through the $\mathbb{S}^{1}$
equator ($y=0$) of the two hemispheres of the $\mathbb{S}^{2}$. Let us now
study the $\rho =\infty $ asymptotic region in the $AdS_3$ foliation. To
this end, we shall consider the boundary change of coordinates 
\begin{eqnarray}
u &=&\frac{1}{2}\ln \left( \frac{2-y^{2}+2\cos \left( \tau -\phi \right) 
\sqrt{1-y^{2}}}{2-y^{2}-2\cos \left( \tau -\phi \right) \sqrt{1-y^{2}}}%
\right) \text{ },  \label{CC1} \\
\sinh \theta &=&\frac{2\sin \left( \phi -\tau \right) \sqrt{1-y^{2}}}{y^{2}}%
\text{ }, \\
\sin t &=&\frac{\sin (2\tau )-\sin (2\phi )\left( 1-y^{2}\right) }{\sqrt{%
\left( 2-y^{2}\right) ^{2}-4\left( 1-y^{2}\right) \cos \left( \tau -\phi
\right) ^{2}}}\text{ }.  \label{CC3}
\end{eqnarray}%
We pick the natural representative of the conformal boundary of $AdS_{4}$ in
the foliation by $AdS_{3}$ space-times (\ref{AdS4}) as given by the metric

\begin{equation}
ds_{3D}^{2}=\frac{1}{\,4}\big[-\cosh ^{2}\theta \,dt^{2}+d\theta ^{2}\big]+%
\frac{1}{\,4}\,(du+\sinh \theta dt)^{2}\,,  \label{AdS3}
\end{equation}%
which is globally $AdS_{3}$ space-time if all the coordinates cover the real
line. If $u\sim u+a$ and the other coordinates cover the whole real line
this three dimensional background is the Coussaert-Henneaux space-time. If
we plug the change of coordinates (\ref{CC1})-(\ref{CC3}) in (\ref{AdS3}) we
get

\begin{equation}
ds_{3D}^{2}=y^{-2}\left( -d\tau ^{2}+\frac{dy^{2}}{1-y^{2}}+\left(
1-y^{2}\right) d\phi ^{2}\right) \text{ .}  \label{3d}
\end{equation}%
Indeed, it can be seen from the boundary point of view that the region of $%
\theta =\pm \infty$ of (\ref{AdS3}) corresponds to $y=0$. This is a surface
with the topology of $S^{1}\times \mathbb{R}$, which is the usual conformal
boundary of $AdS_{3}$ and also the border between the two hemispheres of the
usual conformal boundary of $AdS_{4}$, $\mathbb{S}^{2}\times \mathbb{R}$.

Here we want to remark a very interesting subtlety. As is well known, $AdS$
does not have a boundary but a conformal boundary. The representative of the
conformal boundary given by (\ref{AdS3}) is in a different equivalence class
than $\mathbb{S}^{2}\times \mathbb{R}$. This can be seen directly from (\ref%
{3d}) as the conformal factor that relate these two metrics is singular at $%
y=0$. Therefore, we conclude that the slicing of $AdS_{4}$ by $AdS_{3}$ has
two connected $AdS_{3}$ boundaries if we pick the boundary representative in
the equivalence class of (\ref{AdS3}) and one boundary if we pick the
representative

\begin{equation}
d\bar{s}_{3D}^{2}=\omega (u,\theta )^{2}ds_{3D}^{2}=\frac{\omega (u,\theta
)^{2}}{\,4}\big[-\cosh ^{2}\theta \,dt^{2}+d\theta ^{2}\big]+\frac{\omega
(u,\theta )^{2}}{\,4}\,(du+\sinh \theta dt)^{2}\,,
\end{equation}%
where 
\begin{equation}
\omega (u,\theta )^{2}=\frac{2}{1+\cosh \theta \cosh u}\text{ .}
\end{equation}%
Let us now pass to study a representative of the conformal boundary of the
wormhole space-time (\ref{eq:class-of-metrics}), given by the metric 
\begin{equation}
ds_{\partial M}^{2}=\frac{1}{\,4}\big[-\cosh ^{2}\theta \,dt^{2}+d\theta ^{2}%
\big]+\frac{1}{\,\sigma }\,(du+\sinh \theta dt)^{2}\,,  \nonumber
\label{Warped}
\end{equation}%
which is a quotient of space like warped $AdS_{3}$ when $\sigma \neq 4\,$.
The quotient is product of the identification $u\sim u+a$. If we plug the
change of coordinates (\ref{CC1})-(\ref{CC3}) in (\ref{Warped}) we get 
\begin{eqnarray}
ds_{\partial M}^{2} &=&y^{-2}\Big(-d\tau ^{2}+\frac{dy^{2}}{1-y^{2}}+\left(
1-y^{2}\right) d\phi ^{2}\Big)  \label{boundary} \\
&&+y^{-2}\Big(\frac{4}{\,\sigma }-1\Big)\left( 1-y^{2}\right) \,\Big[\cos
(\tau -\phi )\frac{dy}{1-y^{2}}-y^{-1}\sin \left( \tau -\phi \right) \left(
d\tau +d\phi \right) \Big]^{2}\,.  \nonumber
\end{eqnarray}%
For $\sigma =4$, this is exactly the same boundary than the locally $AdS_{4}$
we just discussed, so we shall consider from now on only the $\sigma \neq 4$
case. The first important aspect of this metric is that $y^{2}ds_{\partial
M}^{2}$ is not locally $\mathbb{S}^{2}\times \mathbb{R}$. Moreover, an
straightforward calculation of the Riemann tensor of (\ref{Warped}) yields 
\begin{equation}
R^{t\theta }{\!}_{t\theta }=4\left( 1-3\sigma ^{-1}\right) \,,\quad R^{tu}{\!%
}_{tu}=R^{\theta u}{\!}_{\theta u}=4\sigma ^{-1}\,,\quad R^{\theta u}{\!}%
_{t\theta }=4\sinh \theta \left( 1-4\sigma ^{-1}\right) \text{ .}
\label{PPS}
\end{equation}%
Hence, from (\ref{PPS}) we observe that warped $AdS_{3}$ develops a
parallelly propagated curvature singularity (PPS)\ at $\theta =\pm \infty $,
we discuss this in more detail in the section on geodesics below. One could
insist in try to construct a space-time with a single boundary by picking a
metric representative in the class of $y^{2}ds_{\partial M}^{2}$. However
this metric has a curvature singularity. An straightforward computation of
the Ricci scalar yields 
\begin{equation}
R(y^{2}ds_{\partial M}^{2})_{\theta =\pm \infty }=2(1-4\sigma
^{-1})e^{\left\vert \theta \right\vert }+O(e^{\left\vert \theta \right\vert
/2})
\end{equation}%
Therefore, the two warped AdS boundaries cannot be joined in a single
compact boundary by a conformal transformation.


\subsection{Geodesics and a parallel propagating singularity}

Our wormhole avoids the use of matter violating the null energy condition
because the throat has a minimal $\mathbb{S}^{1}$ instead of a minimal $%
\mathbb{S}^{2}$ as required in \cite{Hochberg:1998ii}. We would like to
discuss here the different families of null geodesics and use them to
clarify the structure of the conformal boundary just described. A
representative of the conformal boundary of the wormhole space-time (\ref%
{eq:class-of-metrics}) is given by the metric

\begin{equation}
ds_{\partial M}^{2}=\frac{\ell ^{2}}{\,4}\big[-\cosh ^{2}\theta
\,dt^{2}+d\theta ^{2}\big]+\frac{\ell ^{2}}{\,\sigma}\,(du+\sinh \theta
dt)^{2}\,,  \label{space-time}
\end{equation}%
and we set again $\ell =1$. Let us pass to study its geodesics. This
space-time has four Killing vectors \eqref{eq:killing-vectors}, which yield
four conserved charges along geodesic motion 
\begin{equation}
Q_{[i]}=4\dot{x}^{\mu }\xi _{\mu \lbrack i]}  \label{Charges}
\end{equation}%
where $\dot{x}^{\mu }=\left( \dot{t},\dot{\theta},\dot{u}\right) $. These
charges allow to find the velocities 
\begin{eqnarray}
\dot{t} &=&-Q_{1}+L\tanh \theta \sin \left( t-t_{0}\right) \\
\dot{\theta} &=&L\cos \left( t-t_{0}\right)  \label{EQ2} \\
\dot{u} &=&-L\frac{\sin t}{\cosh \theta }+\left( \frac{\lambda ^{2}-1}{%
\lambda ^{2}}\right) \left( Q_{1}\sinh \theta +L\cosh \theta \sin \left(
t-t_{0}\right) \right)
\end{eqnarray}%
where $\lambda^2 =\frac{4}{\sigma },$ $Q_{2}=L\sin t_{0}$ and $Q_{3}=L\cos
t_{0}$ and consistency of (\ref{eq:class-of-metrics}) requires $Q_{1}=Q_{4}$%
. The condition that the geodesic is null yields 
\begin{eqnarray}
0 &=&g_{\mu \nu }\dot{x}^{\mu }\dot{x}^{\nu }  \nonumber \\
&=&\frac{\left( 1-\lambda ^{2}\right) }{4\lambda ^{2}}\left[ \left(
L^{2}\sin ^{2}\left( t-t_{0}\right) +Q_{1}^{2}\right) \cosh ^{2}\theta
+LQ_{1}\sinh 2\theta \sin \left( t-t_{0}\right) \right]  \nonumber \\
&+&\frac{L^{2}\lambda ^{2}-Q_{1}^{2}}{4\lambda ^{2}}  \label{null}
\end{eqnarray}%
plugging (\ref{EQ2}) in (\ref{null}) we obtain an equation in terms of $\dot{%
\theta}$ and functions of $\theta$ only. It has two solutions for $\dot{%
\theta}^{2}$ 
\begin{equation}
\dot{\theta}_{\pm }^{2}=\frac{L^{2}-Q_{1}^{2}}{L^{2}}+\frac{\lambda
^{2}L^{2}+Q_{1}^{2}-2\lambda ^{2}Q_{1}^{2}}{L^{2}\left( 1-\lambda
^{2}\right) \cosh ^{2}\theta }\pm \frac{\sin \theta }{\cosh ^{2}\theta }%
\frac{\lambda Q_{1}}{L^{2}}\sqrt{\frac{Q_{1}^{2}-L^{2}}{1-\lambda ^{2}}}%
\text{ .}  \label{sqrt}
\end{equation}%
Supersymmetric wormholes have $\lambda \leq 1$. The case with $\lambda =1$
is special as it corresponds to locally AdS$_{3}$ space-time and we will
discuss it below. When $\lambda <1$ the integrals of motion must satisfy $%
Q_{1}^{2}-L^{2}\geq 0$ for the square root to be a real number in (\ref{sqrt}%
). This in turn implies that the motion is confined. Indeed, when $%
Q_{1}^{2}-L^{2}\geq 0$ we see that $\dot{\theta}_{\pm }^{2}<0$ for large
enough $\left\vert \theta \right\vert $, which implies that existence of a
turning point at some point when $\dot{\theta}_{\pm }^{2}=0\,$. Hence,
generic null geodesics are confined to live at bounded $\theta $.

There is an exception. Namely, whenever $L^{2}-Q_{1}^{2}=0$. When we replace
the condition $Q_{1}=\pm L$ in \eqref{null} yields 
\begin{equation}
\theta =\pm \frac{1}{2}\ln \left( \frac{1-\sin (t-t_0)}{1+\sin (t-t_0)}%
\right) \text{ ,}
\end{equation}
which allows to integrate the remaining geodesic equation 
\begin{equation}
u=u_{0}\pm \ln \left( \cos (t-t_0)\right) \text{ .}  \label{homothopy}
\end{equation}
We conclude that this geodesic reaches $\theta =\pm \infty $ at $t-t_0=\frac{%
\pi}{2}$. Note that the condition $\lambda =1$ also enforces $%
L^{2}-Q_{1}^{2}=0$ in (\ref{null}). Now, we shall use this geodesic to
construct a parallelly propagated orthonormal frame (PPO). If the geodesic
that reaches null infinity has tangent vector $X^{\mu }$, a set of vierbeine
where one is aligned along $X^{\mu }$ will define such a basis. So we pick $%
\hat{e}_{\mu }^{0}=X_{\mu }$. In this way we construct the Riemann tensor in
this basis 
\begin{equation}
R_{abcd}^{PP}=R^{\mu \nu \delta \alpha }\hat{e}_{a\mu }\hat{e}_{b\nu } \hat{e%
}_{c\delta }\hat{e}_{d\alpha }\,.
\end{equation}%
If the Ricci tensor is singular in the PPO then it implies that the Riemann
is singular which is the definition of a PPS. We will only need to compute
one component of the Ricci tensor in the PPO to check that this is the case: 
\begin{equation}
R_{\mu \nu }X^{\mu }X^{\nu }=\eta ^{ac}R_{a0c0}^{PP}=2L^{2}\sinh ^{2}\left(
\theta \right) \left( \lambda^{2}-1\right) \, .
\end{equation}
Hence, we see that unless $\lambda =1,$ tidal forces diverge when $\theta
\longrightarrow \pm \infty $. Note however that all curvature components are
constant in a static frame 
\begin{eqnarray}
E^{0} =\cosh \theta dt\,, \qquad E^{1} =d\theta \,, \qquad E^{2} =du+\sinh
\theta dt\,,
\end{eqnarray}
since the Riemann tensor for warped AdS is 
\begin{eqnarray}
R_{0202} =3\lambda ^{2}-4 \,, \qquad R_{0303} =-1\,, \qquad R_{2323} =1\,,
\end{eqnarray}
where we recall that we are using a convention where the Riemann tensor of
AdS is constant and positive.

Hence, we conclude that when $\sigma =4$, the boundaries can be connected by
null geodesics. These geodesics can go from one boundary to the other
without problem. Moreover, the time it takes a geodesic to connect the
boundary through the bulk (which has a minimum at $t=2\pi $) is at least the
double it takes through the boundary. Furthermore, it follows from %
\eqref{homothopy} that this null geodesic winds around the \thinspace $u$
coordinate infinite many times. The curves that winds around $u$ are non
contractible. Hence, the geodesic along the boundary is not-homotopic to the
geodesic along the bulk. As expected in a non-simply connected wormhole
space-time.

When $\sigma \neq 4$, it is impossible to connect the different boundaries
with geodesics due to the presence of a PPS. This singularity is a property
of the warped AdS space-time that seems to have passed unnoticed so far in
the literature, see for instance \cite{Bengtsson:2005zj}.


\section{Topological censorship}

There is a set of results that seem to indicate that wormholes can not
exist. In particular in \cite{Galloway:1999br} it is stated the following
theorem: Let $\mathcal{\ M}$ be a globally hyperbolic
space-time-with-boundary with timelike boundary $\mathcal{I}$ that satisfies
averaged null energy condition. Let $\mathcal{I}_{0}$ be a connected
component of $\mathcal{I}$ of $\mathcal{M}$. Furthermore assume that either
(i) $\mathcal{I}_{0}$ admits a compact spacelike cut or (ii) $\mathcal{M}$
satisfies the generic condition. Then $\mathcal{I}_{0}$ cannot communicate
with any other component of $\mathcal{I}$ .

This is particularly relevant to us, as the wormhole seems to have a
boundary with two disconnected components when $\sigma \neq 4$. The
boundary, given by warped $AdS_{3},$ has non-compact spacelike surfaces. So
we need to verify that the generic condition is not satisfied. The generic
condition is the statement that for every timelike or null geodesic $K_{\mu
} $ there is a point where $K_{\left[ \sigma \right. }R_{\left. \mu \right]
\nu \lambda \left[ \alpha \right. }K_{\left. \beta \right] }K^{\nu
}K^{\lambda }\neq 0$.

It is possible to verify that, our wormhole spacetime 
\begin{equation}
ds^{2}=\frac{4\,\ell ^{4}\,dr^{2}}{\sigma ^{2}f(r)}+h(r)\big[-\cosh
^{2}\theta \,dt^{2}+d\theta ^{2}\big]+f(r)(du+\sinh \theta dt)^{2}\,,
\end{equation}
satisfies
\[
K_{\left[ \sigma \right. }R_{\left. \mu \right] \nu \lambda \left[ \alpha
\right. }K_{\left. \beta \right] }K^{\nu }K^{\lambda }=0\,,
\]%
for the following null geodesic 
\[
K=\frac{1}{h(r)\cosh ^{2}\theta }\partial _{t}-\frac{1}{h(r)\cosh \theta }%
\partial _{\theta }-\frac{\sinh \theta }{h(r)\cosh ^{2}\theta }\partial
_{u}\,.
\]%
Note that this is independent of the form of the metric functions. Hence,
our wormhole avoids topological censorship by not being within the
hypothesis of the theorem.


\section{Conclusion}

\label{sec:discussion} \setcounter{equation}{0} 
In this paper we considered supersymmetric transversable wormholes that are
everywhere regular with and without electromagnetic fields. In this respect
these wormholes are crucially different from black holes, which have a
curvature singularity in the interior of the event horizon. The
supersymmetric wormholes preserve half of the supersymmetries. An
interesting fact is that this situation also exists in an $\mathrm{AdS}$
space upon the introduction of a non-contractible cycle. In that case there
exist potentially eight Killing spinors, but only half of them are globally
defined, as was shown in equation \eqref{eq:AdS-Killing-spinors}. The
supersymmetry algebra then implies that the Killing vectors of this
space-time exhibit the same feature, namely that some of them will not be
globally defined. Note, however, that the latter scenario does not involve
electromagnetic fields.

It is worth mentioning that wormhole geometries in asymptotically AdS
space-times have received attention in connection with holography (see e.g. 
\cite{Maldacena:2018lmt}) The presence of multiple boundaries would then
create the possibility of couplings between different CFTs. It had already
been noted earlier that the interaction between two CFTs opens a throat in
the bulk that causally connects the two boundaries \cite{Gao:2016bin}.
However, most of these settings require non-local interactions between the
boundaries for the wormhole throat to open, a feature that is not present in
the construction of this paper.

The supersymmetric, charged, transversable wormholes provide a concrete
physical realization of the Weyl's idea referred to in the introduction.
Non-trivial electromagnetic field lines can be supported by a geometry that
is consistent with the Einstein-Maxwell system.


\section*{Acknowledgements}

We thank Nava Gaddam, Adolfo Guarino, Mario Trigiante and Antoine Van
Proeyen for valuable discussions. We also would like to thank an anonymous
referee who motivate us to include the analysis of sections 5.1 and 5.2. We
wish to thank the project MEC80170073 of CONICYT, Chile, which made this
work possible. We would like to thank the support of Proyecto de cooperaci%
\'{o}n internacional 2019/13231-7 FAPESP/ANID. The research of AA is
supported in part by the Fondecyt Grants 1170279 and 1161418 and the
Alexander von Humboldt Foundation. The research of JO is supported in part
by the Fondecyt Grant 1181047. AA wishes to thank to the warm hospitality at
the Albert Einstein Institut, Golm, Germany and at the Institute for
Theoretical Physics, Utrecht University, Netherlands, where part of this
work was carried out.



\end{document}